\documentclass[conference]{IEEEtran}
\IEEEoverridecommandlockouts
\usepackage{cite}
\usepackage{amsmath,amssymb,amsfonts}
\usepackage{algorithmic}
\usepackage{graphicx}
\usepackage{textcomp}
\usepackage{xcolor}

\usepackage{booktabs} % For formal tables
\usepackage{multirow,tabu}
\usepackage{diagbox}
\usepackage{tikz}
\usetikzlibrary{chains,decorations.pathreplacing,scopes}
\usetikzlibrary{shapes.multipart}
\usepackage[caption=false,font=normalsize,labelfont=sf,textfont=sf]{subfig}
\usetikzlibrary{shapes,arrows,decorations.pathreplacing,calc,patterns,chains}
\tikzstyle{arrow}=[draw, -latex]
\tikzstyle{vertex}=[circle,fill=black!25,minimum size=20pt,inner sep=0pt]
\tikzstyle{selected vertex} = [vertex, fill=red!24]
\tikzstyle{edge} = [draw,thick,-]
\tikzstyle{weight} = [font=\small]
\tikzstyle{selected edge} = [draw,line width=5pt,-,red!50]
\tikzstyle{ignored edge} = [draw,line width=5pt,-,black!20]
\usepackage{xspace}
\usepackage{pgfplots}
\usepackage{xcolor}
\usepackage{url}
\usepackage{float}
\usepackage[inline]{enumitem}
\usepackage[colorinlistoftodos]{todonotes}

\pgfplotsset{compat=newest}
\usepgfplotslibrary{fillbetween}
\usepackage{dblfloatfix} %to enable figures at the bottom of a page
\usepackage{mathtools,boldline,amsthm,amssymb,mathrsfs}

\usepackage{ragged2e, colortbl} %advanced stuff for Table I

\providecommand{\name}{\textnormal{Bifrost}\xspace} %name of our alogrithm, use macro to avoid misspelling 
\providecommand{\bifrost}{\name} %name of our alogrithm, use macro to avoid misspelling
 
\providecommand{\ygg}{Yggdrasil\xspace}

\providecommand{\DD}{DD\xspace}

\providecommand{\user}{\ensuremath{{\sf Client}}\xspace} %client-side algorithm
\providecommand{\users}{\ensuremath{{\sf Clients}}\xspace} %client-side algorithm
\providecommand{\cloud}{\ensuremath{{\sf Cloud}}\xspace} %server-side algorithm
\providecommand{\sender}{\ensuremath{{\sf Sender}}\xspace} %sender
\providecommand{\receiver}{\ensuremath{{\sf Receiver}}\xspace} %receiver

\providecommand{\file}{\ensuremath{F}\xspace} %file / chunck / string
\providecommand{\id}{\ensuremath{{\sf fid}}\xspace} %identifier
\providecommand{\deviation}{\ensuremath{ {C}}\xspace} %deviation on cloud
\providecommand{\local}{\ensuremath{D}\xspace} %data locally stored by clinets
\providecommand{\enclocal}{\ensuremath{E}\xspace} %Encrypted deviation
\providecommand{\hmac}{\ensuremath{M}\xspace} %HMAC of the file
\providecommand{\hmackey}{\ensuremath{K^{\prime}}\xspace}
\providecommand{\enckey}{\ensuremath{K}\xspace}

\providecommand{\HMAC}{\textnormal{\sc{HMAC}}\xspace} %HMAC function
\providecommand{\PRNG}{\textnormal{\sc{PRNG}}\xspace} %PRNG function
\providecommand{\encrypt}{\textnormal{\sc{ENC}}\xspace} %encrypt function
\providecommand{\outsource}{\ensuremath{G}\xspace} %data outsorced to the cloud
\providecommand{\database}{\ensuremath{DB}\xspace} %database
\providecommand{\transformation}{\textnormal{{\sc{TRANSF}}}\xspace}

\providecommand{\bits}{\ensuremath{k}\xspace}  % bits per symbol
\providecommand{\nodel}{\ensuremath{\alpha}\xspace}  % number of deletions
\providecommand{\nof}{\ensuremath{f}\xspace}  % number of records
\providecommand{\sorg}{\ensuremath{s_{\file}}\xspace} %size of each string
\providecommand{\sbase}{\ensuremath{s_{\outsource}}\xspace} %size of each outsource
\providecommand{\sdev}{\ensuremath{s_{d}}\xspace} %size of each deviation
\providecommand{\sizefid}{\ensuremath{s_{\id}}\xspace}  % size of identifier
\providecommand{\norg}{\ensuremath{n_{\file}}\xspace} % symbols per file
\providecommand{\nbase}{\ensuremath{n_{\outsource}}\xspace} % symbols per base

\providecommand{\compsize}{\ensuremath{|size|}\xspace}

\providecommand{\transsize}{\ensuremath{T}\xspace}
\providecommand{\compratio}{\ensuremath{\mathcal{C}}\xspace} %total comp
\providecommand{\adversary}{\ensuremath{\mathcal{A}}\xspace}		
 \newtheorem{assumption}{Assumption}
 
\def\BibTeX{{\rm B\kern-.05em{\sc i\kern-.025em b}\kern-.08em
    T\kern-.1667em\lower.7ex\hbox{E}\kern-.125emX}}
\begin{document}

\title{Bifrost: Secure, Scalable and Efficient File Sharing System Using Dual Deduplication
}
\author{\IEEEauthorblockN{Hadi Sehat$^{1}$, Elena Pagnin$^{2}$, Daniel E. Lucani$^{1}$}%, Claudio Orlandi$^{3}$}
	\IEEEauthorblockA{$^{1}$Agile Cloud Lab, Department of Electrical and Computer Engineering, DIGIT, Aarhus University, Aarhus, Denmark\\
		$^{2}$ Department of Electrical and Information Technology, Lund University, Lund, Sweden\\
		%    $^{3}$ Department of Computer Science, DIGIT, Aarhus University, Aarhus, Denmark\\
		\{hadi,daniel.lucani\}@ece.au.dk,elena.pagnin@eit.lth.se}%, orlandi@cs.au.dk}
	\vspace{-1.5em}
}
%\and
%\and
%\and

\maketitle

\begin{abstract}

We consider the problem of sharing sensitive or valuable files across %multiple 
users while partially relying on a common, untrusted third-party, e.g., a Cloud Storage Provider (CSP).
Although users can rely on a secure peer-to-peer (P2P) channel for file sharing, this introduces potential delay on the data transfer and requires the sender to remain active and connected while the transfer process occurs.
Instead of using the P2P channel for the entire file, users can upload information about the file on a common CSP and share only the essential information that enables the receiver to download and recover the original file.
This paper introduces \name, an innovative file sharing system inspired by recent results on \textit{dual deduplication}.
\name achieves the desired functionality and simultaneously guarantees that
(1) the CSP can efficiently compress outsourced data;
(2) the secure P2P channel is used only to transmit short, but crucial information;
(3) users can check for data integrity, i.e., detect if the CSP alters the outsourced data; and
(4) only the sender (data owner) and the intended receiver can access the file after sharing,
 i.e., the cloud or no malicious adversary can infer useful information about the shared file.
%We achieve this by leveraging a recent technique for secure file outsourcing called dual deduplication (Sehat et al. in ICC 2021), and basic cryptographic primitives such as HMAC and symmetric encryption.
%Taking inspiration from recent works for secure file outsourcing using \textit{dual deduplication},
We analyze compression and bandwidth performance using a proof-of-concept implementation.
Our experiments show that secure file sharing can be achieved by sending only $650$~bits on the P2P channel, irrespective of file size, while the CSP that aids the sharing can enjoy a compression rate of $86.9$~\%.
%We develop a proof-of-concept implementation of our system and analyze its performance in terms of compression potential for the CSP and bandwidth requirements on the P2P channel.
%Our experiments show that we achieve secure file sharing sending only 650 bits on the P2P channel, regardless of the size of the data,
%and the CSP aiding in the file sharing enjoys a
%compression rate of 78.2\%.
\end{abstract}

%
%\begin{IEEEkeywords}
%Data Compression, Data Privacy, Deduplication, Generalized Deduplication.
%\end{IEEEkeywords}

% To allow compilation of the file
% !TeX root = ./../Bifrost.tex

\section{Introduction}\label{intro}
Secure file sharing between multiple devices has been a major research topic in recent years.
Existing approaches tend to offer efficient solutions, especially in settings where the sharing process is aided by a cloud storage system.
However, this setting inherently raises a number of privacy and security concerns on users' outsourced data.
In particular, if no protection mechanism is put in place, users' data is leaked to the cloud, to an adversary that gains access to the cloud, or to malicious actors monitoring the transmission channel.
A traditional method to mitigate such threats is to encrypt the data, upload the ciphertext to the cloud, and send the key required to decrypt the data using a private secret channel between the two users.
%However, this method is vulnerable if the encryption can be broken by an adversary.
%\elena{the fact that the encryption scheme can be broken applies to any solution, also to Bifrost, so it is not a good argument}
%In addition,
While this simple solution works for users' privacy, it affects the performance of the cloud aiding in the file sharing procedure.
Handling ciphertext rather than plaintext data prevents the cloud from efficiently store data. Intuitively, this is due to the fact that encrypted data is designed to look uniformly random, thus, preventing any compression potential.

\begin{figure*}[!b]
	\centering
	\begin{tikzpicture}[
		node distance=0pt,
		%start chain = A, %going right,
		X/.style = {align=left, rectangle, draw,% styles of nodes in string (chain)
			minimum height=4ex,
			outer sep=0pt},
		]
		%File
		\node (file) at (-0.65,0) [X, minimum width = 35.3ex] {\file};
		
		%chunks
		\node (chunks) at (5,0.35) [X, minimum width = 8.8ex] {A};
		\node (chunks2) at (6.7,0.35) [X, minimum width = 8.8ex] {B};
		\node (chunks3) at (5,-0.35) [X, minimum width = 8.8ex] {C};
		\node at (6.7,-0.35) [X, minimum width = 8.8ex] {D};
		
		%gen-chunks
		\node at (11,0.35) [X, minimum width = 7.8ex] {$\mathcal{A}$};
		\node at (12.1,0.35) [X, minimum width = 1ex] {$\Delta_{A}$};
		\node (genchunks) at (11,-0.35) [X, minimum width = 7.8ex] {$\mathcal{C}$};
		\node at (12.1,-0.35) [X, minimum width = 1ex] {$\Delta_{C}$};
		
		\node (genchunks2) at (13.25,0.35) [X, minimum width = 7.8ex] {$\mathcal{B}$};
		\node at (14.35,0.35) [X, minimum width = 1ex] {$\Delta_{B}$};
		\node at (13.25,-0.35) [X, minimum width = 7.8ex] {$\mathcal{D}$};
		\node at (14.35,-0.35) [X, minimum width = 1ex] {$\Delta_{D}$};
		
		%outsource to cloud
		\node (cl12) at (9.5,3.45) [X, minimum width = 7.8ex] {$\mathcal{A}$};
		\node at (9.5,2.75) [X, minimum width = 7.8ex] {$\mathcal{C}$};	
		\node (cl11) at (10.85,3.45) [X, minimum width = 7.8ex] {$\mathcal{B}$};
		\node at (10.85,2.75) [X, minimum width = 7.8ex] {$\mathcal{D}$};
		
		%deduplication on cloud
		\node (cl22) at (3.8,3.45) [X, minimum width = 6.8ex] {$\mathsf{A}$};
		\node at (4.8,3.45) [X, minimum width = 1ex] {$\delta_{A}$};
		\node at (3.8,2.75) [X, minimum width = 6.8ex] {$\mathsf{A}$};
		\node at (4.8,2.75) [X, minimum width = 1ex] {$\delta_{C}$};
		
		\node at (5.95,3.45) [X, minimum width = 7.8ex] {$\mathsf{B}$};
		\node (cl21) at (6.95,3.45) [X, minimum width = 1ex] {$\delta_{B}$};
		\node at (5.95,2.75) [X, minimum width = 7.8ex] {$\mathsf{B}$};
		\node at (6.95,2.75) [X, minimum width = 1ex] {$\delta_{D}$};
		
		%transformation on cloud
		\node at (-1.75,3.45) [X, minimum width = 6.8ex] {$\mathsf{A}$};
		\node at (-0.8,3.45) [X, minimum width = 1ex] {$\delta_{A}$};
		%\node at (0.8,4.75) [X, minimum width = 6.8ex] {$\mathsf{A}$};
		\node at (-0.8,2.75) [X, minimum width = 1ex] {$\delta_{C}$};
		
		\node at (0.2,3.45) [X, minimum width = 7.8ex] {$\mathsf{B}$};
		\node (cl31) at (1.2,3.45) [X, minimum width = 1ex] {$\delta_{B}$};
		%	\node at (0.95,4.75) [X, minimum width = 7.8ex] {$\mathsf{B}$};
		\node at (-0.1,2.75) [X, minimum width = 1ex] {$\delta_{D}$};

		\node[ cloud, cloud puffs=19, cloud puff arc=50, minimum width=18cm, minimum height=3cm, align=center, draw] at (5, 3) {};
		\draw (-4,0.75) rectangle (15,-0.75);
		\node at (-3,1) {\textbf{Client}};
		\node at (-3,3.9) {\textbf{Cloud}};
		\draw [arrow] (file.east) -- ([yshift = -10pt] chunks.west) node[midway,above] {chunking};
		\draw [arrow] ([yshift = -10pt]chunks2.east) -- ([yshift = 10pt] genchunks.west) node[midway,above] {transformation};
		\path [arrow]  ([xshift=-20pt]genchunks2.north) |-  ([xshift = 20 pt,yshift = -10pt] cl11.east)  -- ([yshift = -10pt] cl11.east);
		\draw [arrow] ([yshift = -10pt]cl12.west) -- ([yshift = -10pt] cl21.east) node[midway,above] {transf};
		\draw [arrow] ([yshift = -10pt]cl22.west) -- ([yshift = -10pt] cl31.east) node[midway,above] {transf};
		\node at(13.2,1.1) {upload};
	\end{tikzpicture}
	\caption{An illustration of a dual deduplication system.}
	\label{fig:DD}
\end{figure*}
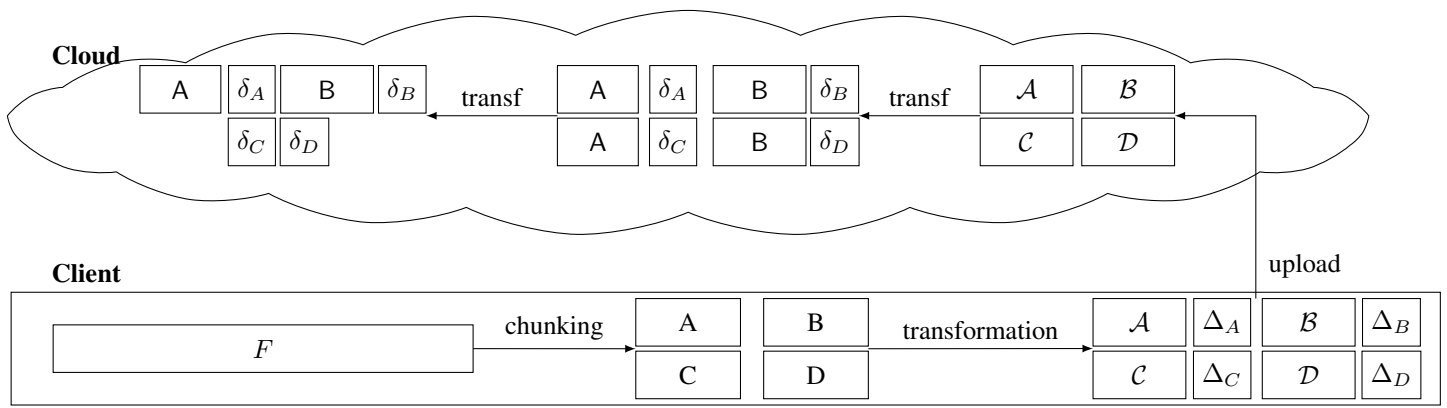

In this work, we propose a secure file sharing system that relies on a Dual Deduplication (\DD) mechanism.
In brief, the authors of~\cite{sehat2021yggdrasil} developed the concept of \DD, which
%generates information-theoretically secure data that can not be decrypted by the adversary,
%while allowing efficient compression potential on the cloud.
allows users to transform their input data into a compression-friendly data fragment, and a short recovery fragment that should be kept private by the user.
The compression-friendly piece is outsourced to the cloud,
and protects the privacy of the user by introducing uncertainty on the original data.
This uncertainty is produced by using information-theoretical mechanisms to hide information
in the process of generating the outsourced data.
% and is protected using information-theoretical mechanisms to hide information about the original data and create uncertainty on the potential sources of this fragment.
%The secret recovery piece is stored locally by the user.
%In \cite{sehat2021yggdrasil}, Sehat et al. develop a dual deduplication mechanism called \ygg, inspired by the tree of life in the Norse Mythology.
The protocol was named \ygg~\cite{sehat2021yggdrasil}.
Our approach is inspired by \ygg and aims at transferring information among clients.
We name our system \bifrost, like the burning rainbow bridge that connects Midgard (Earth) and Asgard (the realm of the Gods).
Being a secure \DD mechanism, \ygg focuses on balancing data privacy and data compression.
However, it does not guarantee data integrity, a central feature in file sharing.
%We further articulate on a specific \DD mechanism called \ygg in Section \ref{sec:yggrdasil}.
%\bifrost address this by including an integrity check mechanism.
%In a nutshell, in addition to the transformations performed by \ygg,
%\bifrost employs a Hash-Based Message Authentication Code (HMAC) to generate an integrity tag on the original input data.
%This mitigates the effect of a malicious CSP trying to alter outsourced data, and thus preventing the receiving party from recovering the correct file.

\begin{figure}[!t]
	\centering
	\begin{tikzpicture}[
		node distance=0pt,
		%start chain = A, %going right,
		X/.style = {align=left, rectangle, draw,% styles of nodes in string (chain)
			minimum height=4ex,
			outer sep=0pt},
		]
		%File
		\node (file) at (5.6,2.4) [X, minimum width = 20ex, minimum height=8ex] {\file};
		\node (file2) at (9.75,2.4) [X, minimum width = 20ex, minimum height=8ex] {\file};
		
		%chunks
		\node (chunks) at (5,0.35) [X, minimum width = 8.8ex] {A};
		\node (chunks2) at (6.55,0.35) [X, minimum width = 8.8ex] {B};
		\node (chunks3) at (5,-0.35) [X, minimum width = 8.8ex] {C};
		\node at (6.55,-0.35) [X, minimum width = 8.8ex] {A};
		
		%gen-chunks
		\node (chunks2)at (8.25,0.35) [X, minimum width = 7.8ex] {$\mathcal{A}$};
		\node at (9.35,0.35) [X, minimum width = 1ex] {$\Delta_{A}$};
		\node (genchunks) at (8.25,-0.35) [X, minimum width = 7.8ex] {$\mathcal{B}$};
		\node at (9.35,-0.35) [X, minimum width = 1ex] {$\Delta_{C}$};
		
		\node at (10.5,0.35) [X, minimum width = 7.8ex] {$\mathcal{B}$};
		\node at (11.6,0.35) [X, minimum width = 1ex] {$\Delta_{B}$};
		\node at (10.5,-0.35) [X, minimum width = 7.8ex] {$\mathcal{A}$};
		\node at (11.6,-0.35) [X, minimum width = 1ex] {$\Delta_{A}$};
		
		%dedup
		\node (dedup) at (5,-2) [X, minimum width = 8.8ex] {A};
		\node  at (6.55,-2) [X, minimum width = 8.8ex] {B};
		\node at (5.85,-2.7) [X, minimum width = 8.8ex] {C};
		%		\node (file) at (2.8,-2.7) [X, minimum width = 4ex] {A};
		
		%gen-dedup
		\node  (gendedup) at (8.25,-2) [X, minimum width = 7.8ex] {$\mathcal{A}$};
		\node  at (9.35,-2) [X, minimum width = 1ex] {$\Delta_{A}$};
		\node at (8.25,-2.7) [X, minimum width = 7.8ex] {$\mathcal{B}$};
		\node at (9.35,-2.7) [X, minimum width = 1ex] {$\Delta_{C}$};
		
		%		\node (file) at (6.25,-2) [X, minimum width = 3.8ex] {$\mathcal{B}$};
		\node  at (10.25,-2.35) [X, minimum width = 1ex] {$\Delta_{B}$};
		%		\node (file) at (6.25,-2.7) [X, minimum width = 3.8ex] {$\mathcal{A}$};
		%		\node (file) at (7,-2.7) [X, minimum width = 1ex] {$\Delta_{A}$};
		
		\draw [->] ([xshift=-0.3pt]file.south) -- ([xshift = 22pt] chunks.north) node[midway,left] {chunk};
		\draw [->] ([xshift=-20.5pt]file2.south) -- ([xshift = 22pt] chunks2.north) node[midway,right] {chunk and transform};
		\draw [->] ([xshift = 22pt]chunks3.south) -- ([xshift = 22pt] dedup.north) node[midway,left] {Deduplication};
		\draw [->] ([xshift = 22pt]genchunks.south) -- ([xshift = 22pt] gendedup.north) node[midway,right] {Gen. Deduplication};
		\draw [-] (7.4,3.2) -- (7.4,-3.2);
	\end{tikzpicture}
	\caption{An illustration of Deduplication and Generalized Deduplication in a storage system.}
	\label{fig:dedups}
\end{figure}
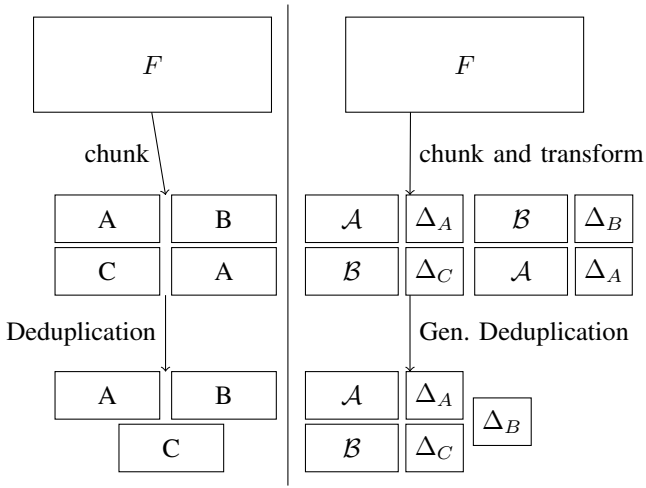

\subsection{Data Deduplication and \ygg}

Before we discuss \name and our contributions,
we take a look into the aspect of data deduplication and how it is implemented in 
\ygg to reduce the fingerprint of data on a cloud storage system.
%Data deduplication~\cite{}, and generalized deduplication~\cite{vestergaard2019generalized} are techniques used to
Data deduplication was introduced in~\cite{welch1984technique} as a technique to 
reduce the data fingerprint by removing the identical chunks of data.
In a deduplication-based storage system, a file is divided into multiple chunks of a pre-defined fixed length.
When the storage identifies a duplicate chunk among the chunks of files in the data,
it does not store the chunk as a whole, but a pointer to its duplicate counterpart.   
These techniques work on redundancy between chunks as opposed to compression that works on redundancy within a
chunk~\cite{constantinescu2011mixing}.
Fig.~\ref{fig:dedups} shows an illustrative example of Deduplication, where a file \file
is divided into four chunks $A$,$B$,$C$ and $A$,
in this case, the deduplication algorithm remove the duplicate chunk $A$.

Although data deduplication is an effective method to reduce the fingerprint of the data,
it can not identify and remove similar chunks from the data. 
Therefore, an alternative approach was proposed in~\cite{vestergaard2019generalized}, 
called Generalized Deduplication (GD). In this method, the chunks are transformed using a transformation technique
into a base and a small deviation pair. The base is used for deduplication while the deviation indicates the difference
between the original chunk and the generated base.
This transformation is performed in a way to ensure mapping of multiple chunks into a single base,
improving the deduplication potential, while having a small deviation that has a low impact on the
storage requirements. In the current state-of-the-art GD implementations, the used transformation techniques include
Hamming codes~\cite{vestergaard2019generalized}, Reed-Solomon Codes~\cite{nielsen2019alexandria} and edit operations such as deletion and swap~\cite{sehat2021yggdrasil}. 
Fig.~\ref{fig:dedups} shows a high level example of GD on the same file with the same set of chunks, 
where both chunks $B$ and $C$ are transformed into the same base $\mathcal{B}$.

DD uses the aspect of GD to provide privacy for a client that wants to store data
on a cloud storage system that performs deduplication. The idea behind DD is the fact that without the exact knowledge on
the deviation, the original chunk cannot be constructed directly from the base.
Therefore, in DD, the client performs the transformation, creating base and deviation, and sends the base to the cloud,
while storing the deviation locally, hiding its components from the cloud.
In general, in DD, both client and cloud engage by applying transformations on the data.
The goal of the client is to provide privacy by creating a small deviation that is stored locally,
and the goal of the cloud is to apply deduplication on the received data from the clients.
In \ygg in particular, the transformation that is performed on the client is deletion of symbols.
The idea behind this transformation comes from the aspect of deletion channel~\cite{mitzenmacher2009},
which is proven to be a hard problem.
As a general overview, the client in \ygg deletes some elements from each chunk,
creating a base with a size less than the original chunk, and storing the position and value of the deleted elements
in the deviation. The base is sent to the cloud, where other edit operations, namely swap and change value,
are used for generalized deduplication of the received bases.
More details about \ygg and its performance can be found in~\cite{sehat2021yggdrasil}.
Fig.~\ref{fig:DD} shows a high level view of the
transformations performed in a dual deduplication system.

\subsection{Contribution}
Our main contribution is the design and development the first secure file sharing system based on dual deduplication.
Concretely, we build on \ygg~\cite{sehat2021yggdrasil} and construct \bifrost, a scalable and efficient solution for P2P, server-aided secure file sharing.
In addition to the transformations performed by \ygg,
\bifrost employs a Hash-Based Message Authentication Code (HMAC)
to generate an integrity tag on the original input data.
This HMAC is used to authenticate the file, and
mitigates the effect of 
an untrusted CSP 
altering the outsourced data and preventing the receiving party from recovering the correct file.
To reduce the amount of data required to be sent using the secure P2P channel,
the client encrypts the local secret piece using a symmetric encryption scheme
and sends the encrypted secret piece to the cloud.
This ensures that as long as the encryption scheme is secure,
the cloud cannot learn useful information about the original data,
while it can perform efficient compression on a major portion of the data, 
which is generated by using \ygg.
To share a file with another user or device, the data owner needs only to share the HMAC key used to authenticate the file,
%the file identifier (for retrieval from the cloud), 
and the secret keys required to decrypt the HMAC and the secret piece.% generated by the dual deduplication mechanism.

More specifically, our contributions are the following:
\begin{itemize}
	\item We formalize a system model and a threat model for file sharing using the \DD paradigm.

\item  We design \bifrost, our realization of a secure, scalable and efficient file sharing system using \DD.

\item We develop a proof-of-concept implementation of \bifrost and analyze its performance in terms of compression rate,
provided privacy and required transmission in the P2P channel.

%- We also provide a version optimized for low-storage user environments, called \encbifrost.
%In brief, \encbifrost works as \bifrost except that the short secret piece generated by the dual deduplication mechanism is not stored by the client, but encrypted (using a secret key) and sent as a ciphertext to the cloud.
%Obviously, when sharing a file the secret encryption key is sent in place of the  secret piece.

\item We demonstrate experimentally that \bifrost realizes secure file sharing between users using a cloud storage by sending only $650$ bits, regardless of the size of the data. Moreover, we show that \bifrost achieves a compression rate of 86.9\%, while providing computational security against an untrusted cloud.
\end{itemize}

In the following, we first take a look at some related work in terms of both
secure file sharing in an untrusted environment and DD in Section~\ref{sec:related},
then we introduce the system model and performance metrics in Section~\ref{sec:model}.
We follow by introducing \name and a step-by-step illustration of the protocol in Section~\ref{sec:contribution}.
We then analyze the privacy of \name in Section~\ref{sec:privacyanalysis}, followed by simulation results in
Section~\ref{sec:results}. Finally, we conclude in Section~\ref{sec:conclusion}.  
\section{Related Work}\label{sec:related}
As \name is a secure file sharing system using an untrusted cloud,
and is proposed based on DD, it is natural to look at
the current literature about both these topics.
As far as our knowledge, there is no work that has combined the aspects of
secure file sharing and DD until now.
Therefore, in this Section, we look at some of the state-of-the-art
research on the topics that are the main focus point of \name.

\subsection{Secure File Sharing}

Secure file sharing using an untrusted cloud has been a long-known research topic, especially as providing
secure file sharing would introduce a large amount of overhead to the cloud, the clients or both~\cite{plutus}.
%Therefore, many works have focused on providing scalability, among other performance metrics.
One of the earliest scalable file sharing systems using an untrusted server was Plutus~\cite{plutus},
which proposed a scalable file sharing system in the presence of an untrusted cloud.
Since then, this research area has focused on multiple aspects of secure file sharing,
such as dynamic permission and revocation of read/write rights~\cite{goh2003sirius},
group-based permissions and access to encrypted files~\cite{contiu2019anonymous},
reducing the overhead on the cloud by fixing the number of ciphertexts for each file~\cite{boneh2005collusion},
and allowing deduplication on the ciphertexts~\cite{li2016rekeying}.
As a noteworthy recent research output, SeGShare~\cite{fuhry2020segshare} 
uses enclaves and trusted execution environments to achieve promising results
in terms of scalability and throughput.

The closest related work in this area are the research focused on providing deduplication
on the cloud. However, the major research output 
on this topic~\cite{li2016rekeying,fuhry2020segshare, contiu2019anonymous, qin2017design},
use deterministic content-derived encryption
% as their encryption method,
which uses keys derived from the content in order to encrypt the data.
Although this method has provable security at the moment,
% for the time being, 
the security assumptions for these algorithms might not hold in the future,
where the security of these methods will become questionable.
In contrast, in \name, we argue for privacy against an untrusted cloud while using any form of
encryption on the client. Therefore, \name can be implemented with any encryption method,
depending on the use case and desired privacy.
%is still a valid file sharing system by using a stronger encryption method.

\subsection{Security for Deduplication}
Implementing deduplication for untrusted cloud storage or cloud sharing system
has been studied as providing privacy alongside deduplication.
There have been multiple works in this area, 
all of which can be categorized under one of the three approaches of 
convergence encryption~\cite{whiting1998system},
multi-key revealing encryption (MKRE)~\cite{lucani2020secure},
and DD~\cite{sehat2021yggdrasil}.
Convergence encryption is a message leaked encryption~\cite{akhila2016study} method 
that ensures privacy against untrusted cloud while providing deduplication on identical chunks.
Convergence Encryption can be seen as a random Oracle with consistent output, 
i.e., identical plaintexts always produce identical ciphertexts~\cite{xia2016comprehensive}.
MKRE is an encryption method to provide generalized deduplication on the cloud while providing computational privacy 
proven in the form of random oracle.
DD is a relatively new technique proposed to provide information-theoretic privacy against
an honest cloud, while allowing for deduplication or generalized deduplication to be performed on the data on the cloud.
In DD, instead of encrypting the file, the client uses information-theoretic tools,
such as deletion to introduce uncertainty for an adversary that wishes to reconstruct the original data of the clients.
%More recently, the authors of~\cite{bonsai} proposed Bonsai, which provides a compression technique based on Dual Deduplication.
%In~\cite{}, the cloud uses the statistical properties injected into the outsourced data in order to perform
%deduplication and compression at the same time on the data.
%We use Bonsai as the backbone of \name, to propose a secure file sharing while having an untrusted cloud storage.

% To allow compilation of the file
% !TeX root = ./../Bifrost.tex

\section{System Model and Performance Metrics}\label{sec:model}

Our system consists of a cloud storage, referred to as \cloud,
and multiple devices, referred to as \users.
\users have data that they outsource to the \cloud in order to
both reduce storage demands on their end and enable secure and lightweight file sharing among clients.
In this work, \emph{data} refers to a collection of files.
Each file \file is viewed as a string of \sorg bits,
consisting of \norg symbols of \bits size each.
Obviously, we have $\sorg = \norg\cdot \bits$.
After performing transformations in the \user,
we generate an outsource piece \outsource and a secret piece \local.
\outsource is a string consisting of \nbase symbols, where $\nbase < \norg$,
and therefore has a size of $\sbase = \nbase\cdot\bits$.
\local is the secret piece that contains the information about the
transformations performed in the client.
\local has a size of \sdev.
We assume that the channel between \users and \cloud  as well as the P2P channel between \users are error-free.
We assume that the P2P channel cannot be infiltrated by the adversary.

The adversary \adversary in our system is an untrusted \cloud, and
any malicious \user that is able to access the data stored in the \cloud.
We aim to provide privacy against this adversary in two ways:
\begin{enumerate*}
	\item Using transformations on the \user data, which result in a high uncertainty for the adversary on what the original data was, providing privacy for the data.
	\item Mitigating the possibility of changing the data stored on the \cloud by the adversary.
\end{enumerate*}
We assume that both \users that engage in the file sharing protocol are honest,
and therefore follow the protocol exactly.
We also note that we want to achieve privacy for \user's data 
while providing sufficient compression capabilities for the \cloud.

We analyze the performance of our proposed method (\name) using three different metrics:
\begin{enumerate*}
	\item The compression rate on the \cloud.
	\item The required transmission in the P2P channel.
	\item The privacy against an untrusted \cloud.
\end{enumerate*}

The compression ratio on the \cloud denotes the required size to store the received data
after compression, 
%on the \cloud 
compared to the size of the original data.
The size of the original data, i.e., \database is denoted by $|\database|$.
In order to define the compression ratio, we need to define two parameters:
\begin{enumerate*}
	\item \compsize denotes the required size to store all the received \outsource and \enclocal in the \cloud after compression.
	\item \sizefid denotes the size of the file identifier for each set of $(\outsource, \enclocal)$ sent to the \cloud.
\end{enumerate*}

The compression ratio on the \cloud is then:
\begin{description}
	\item[Compression ratio:]
	\begin{equation}\label{eq:compratio}
		\compratio = \frac{\compsize + \nof\cdot\sizefid}{|\database|},
	\end{equation}
\end{description}
where \nof is the number of files in \database.

The required transmission rate is equal to the size of the data that the \users
need to share in the secure P2P channel.
As mentioned, this information consists of three components.
\begin{enumerate*}
	\item The HMAC of file $\hmac(\file)$, that acts also as the file identifier and has the size of \sizefid.
	\item The secret key required to generate HMAC from file, i.e., \hmackey.
	\item The secret key required to decrypt \enclocal, i.e., \enckey.
\end{enumerate*}

Therefore, the total amount of required transmission for each file is equal to
\begin{description}
	\item[Compression ratio:]
	\begin{equation}\label{eq:transsize}
		\transsize(\file) = \sizefid + |\hmackey| + |\enckey|,
	\end{equation}
\end{description}
 where $|x|$ denote the size of component $x$.

For the privacy of \name, we use the notion of 
random oracle model in order to prove the privacy of \name.
We argue that in \name, a computationally bounded untrusted \cloud doe not gain any
information about the original file of \users.
We prove the privacy of \name in full details
% We argue that the privacy of \name is based on the privacy characteristics of the techniques used to
% upload data to the \cloud and share data. We prove the computational privacy of \name using a random oracle model
in Section~\ref{sec:privacyanalysis}.

% To allow compilation of the file
% !TeX root = ./../Bifrost.tex
\section{\name: Our Secure File Sharing Solution}\label{sec:contribution}

\begin{figure}[!t]
	\includegraphics[scale=.19]{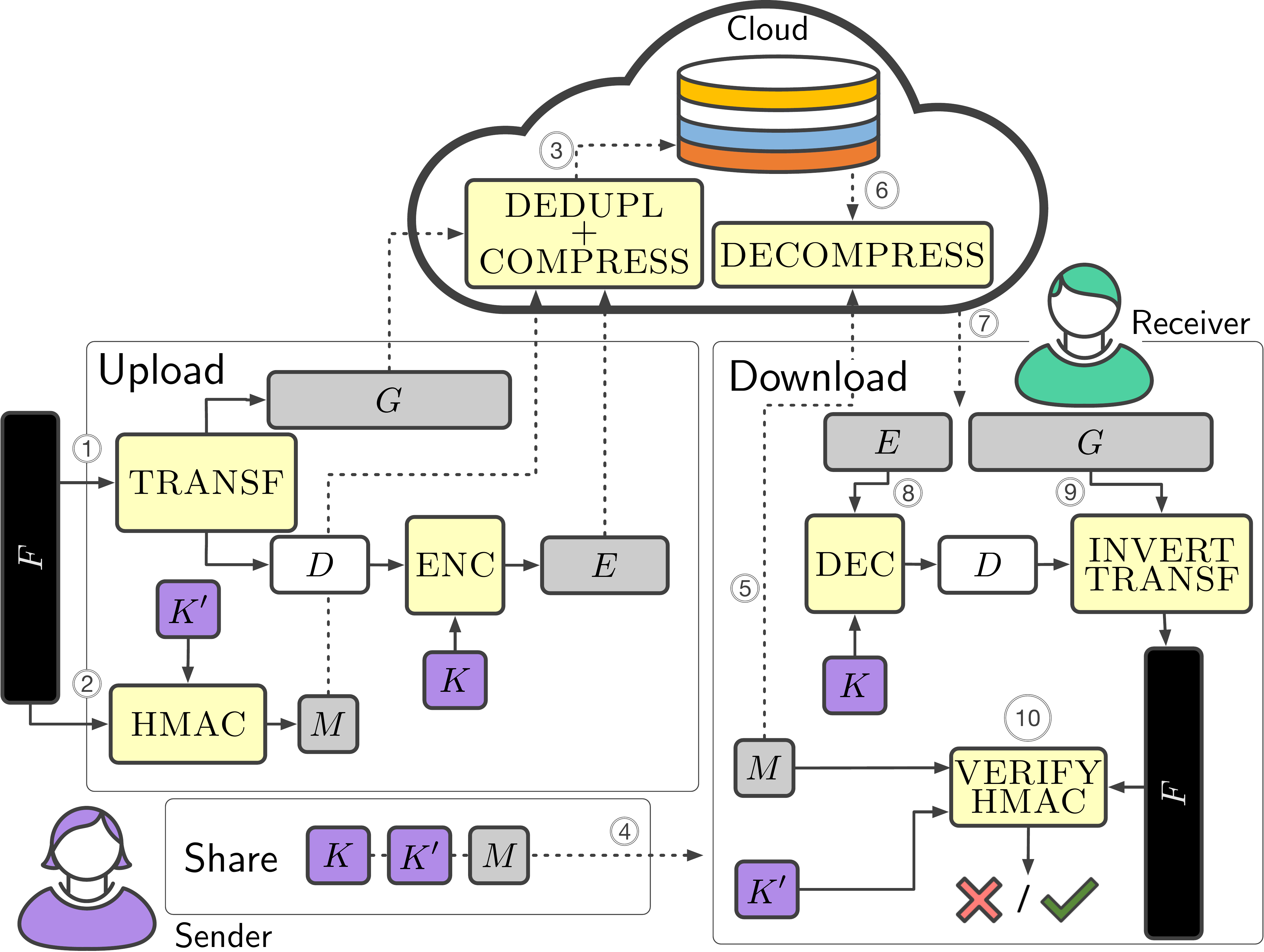}
	\caption{A graphic representation of the information flow and the main procedures in \bifrost.}
	\label{Fig:Bflow}
\end{figure}

In this Section, we describe \name.
First, we discuss how the data is uploaded from the \user to the \cloud,
and then we illustrate the transmissions required between two \users.
We refer to the \user that wants to share its data as \sender, and the receiving \user as \receiver.
Fig.~\ref{Fig:Bflow} shows an illustration of the steps taken by each party,
and the communications between different parties of the system.
We go through these steps in two subsections,
first on uploading data from the \sender to the \cloud
and then on the sharing of information from \sender to \receiver and downloading the data from the \cloud by the \receiver.

%\begin{figure}[!t]
%	\includegraphics[scale=.25]{figure-sources/Bifrost-flow.pdf}
%	\caption{A graphic representation of the information flow and the main procedures in \bifrost.}
%	\label{Fig:Bflow}
%\end{figure}
\subsection{Uploading data from \user to \cloud}

\begin{figure*}[t]
	\centering
	\begin{tikzpicture}[
		node distance=0pt,
		%start chain = A, %going right,
		X/.style = {align=left, rectangle, draw,% styles of nodes in string (chain)
			minimum width=2.5ex, minimum height=2.8ex,
			outer sep=0pt},
		Y/.style = {align=left, rectangle, draw,% styles of nodes in string (chain)
			minimum width=3.5ex, minimum height=3.5ex,
			outer sep=0pt},
		]

		%File
		{
			[start chain =A],
			\node[X,right, on chain = A] {\footnotesize 4};
			\node[X,right, on chain = A] {\footnotesize 1};
			\node[X,right, on chain = A] {\footnotesize 10};
			\node[X,right, on chain = A] {\footnotesize \textcolor{red!80}{11}};
			\node[X,right, on chain = A] {\footnotesize 8};
			\node[X,right, on chain = A] {\footnotesize \textcolor{red!80}{6}};
			\foreach \i in {1,9,14}
			\node[X,right, on chain = A] {\footnotesize \i};
			\node(endA) [X, right, on chain = A] {\footnotesize 15};
		}
		\foreach \x [count=\i] in {0,1,2}
		\node [below=of A-\i]{\textcolor{black!50}{\tiny \x}};
		\node [below=of A-4]{\tiny \textcolor{blue!80}{3}};
		\node [below=of A-5]{\tiny \textcolor{black!50}{4}};
		\node [below=of A-6]{\tiny \textcolor{blue!80}{5}};
		\node [below=of A-7]{\tiny \textcolor{black!50}{6}};
		\node [below=of A-8]{\tiny \textcolor{black!50}{7}};
		\node [below=of A-9]{\tiny \textcolor{black!50}{8}};
		\node [below=of A-10]{\tiny \textcolor{black!50}{9}};

		\node [inner sep=1pt,above=of A-6.north east] {\footnotesize $\file$};

		%PRNG

				{
			[start chain = B],
			\node (sResult1) [X,right, on chain = B] at (3,-1.2) {\tiny \textcolor{blue!80}{5}};
			\node[X,right, on chain = B] {\tiny \textcolor{blue!80}{3}};
		}
		\node (PRNG1) [X, left=0.5cm of B-1, label={\tiny PRNG}, fill=black!30] {};
		\node (s1) [left=1.2cm of B-1]{$s_{1}$};

		\node [below=of B-1]{\textcolor{black!50}{\tiny $P1$}};
		\node [below=of B-2]{\textcolor{black!50}{\tiny $P2$}};

		\draw [arrow] (s1)  -- (PRNG1);
		\draw [arrow] (PRNG1) -- (sResult1);

		%Outsource
		{
			[start chain =A1],
			\node(beginA1)[X,right, on chain = A1] at (6,0) {\footnotesize 4};
			\foreach \i in {1,10,8,1,9,14}
			\node[X,right, on chain = A1] {\footnotesize \i};
			\node (endA1) [X,right, on chain = A1] {\footnotesize 15};
		}

		\foreach \x [count=\i] in {0, ...,7}
		\node [below=of A1-\i]{\textcolor{black!50}{\tiny \x}};

		\node [inner sep=1pt,above=of A1-5.north west] {\footnotesize $\outsource$};

		\path [->] (endA) edge (beginA1);

			%Local
		{
			[start chain =A1D],
			\node(beginA1D)[X,right, on chain = A1D] at (10,0) {\small $s_{1}$};
%			\node[X,right, on chain = A1D] {\small $s_{1}$};
			\node[X,right, on chain = A1D] {\footnotesize \textcolor{red!80}{6}};
			\node[X,right, on chain = A1D] {\footnotesize \textcolor{red!80}{11}};
%			\node[X,right, on chain = A1D] {\footnotesize \textcolor{red!80}{1}};
		}

%		\node [below= of A1D-1]{\textcolor{black!50}{\tiny \invertbit}};
	\node [below=of A1D-2]{\textcolor{black!50}{\tiny  $V1$}};
	\node [below=of A1D-3]{\textcolor{black!50}{\tiny  $V2$}};
%	\node [below=of A1D-5]{\textcolor{black!50}{\tiny  $V3$}};

	\node [inner sep=1pt,above=of A1D-3.north] {\footnotesize Deviation ($local_{1}$)};

	\end{tikzpicture}
	\caption{An illustration of the transformation \transformation in the \user to create \outsource and \local.}
	\label{fig:trf}
\end{figure*}
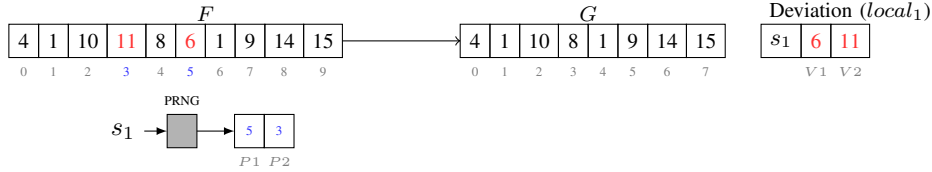

This stage consists of actions from two entities, the \user and the \cloud.
When a \user wants to store a file \file in the \cloud,
(1) it uses the a set of transformations \transformation to create an outsourced data \outsource
and a deviation \local .
This transformation \transformation includes deletion of symbols from the file \file.
These deletions are performed using a Pseudo-Random Number Generator (PRNG),
and mimics the function of a deletion channel to provide privacy for the \user's data~\cite{sehat2021yggdrasil}.
The deleted symbols are identified by using a seeded PRNG,
which outputs $\alpha = \sorg - \sbase$ random numbers between 0 and $\sorg-1$,
pointing to the position of the symbols in \file that are to be deleted.
To give an intuition, Fig.~\ref{fig:trf} shows an example of \transformation
performed on a file \file consisting of 10 symbols of 4 bits each.
In this particular example, The \user deletes 2 symbols.
The numbers generated by the seeded PRNG are 5 and 3.
The \user deletes the symbols in positions 5 and 3, which are 11 and 2,
Generating \outsource; as well as  storing the seed used to create the positions,
alongside the value of the deleted symbols in \local.
The \user is able to regenerate \file from \outsource and \local by creating the
position of the stored value using the same PRNG and the random seed stored in \local,
and simply inserting the values into their respective positions.

(2) The \user then computes the HMAC of the \file using a \HMAC function
with a secret key \hmackey
to generate $\hmac = \HMAC(\file)$.
This value is used as the file identifier in \name, as is sent to the \cloud,
as well as stored locally.
The purpose of \hmac is to provide data integrity against a malicious cloud.
Any authentication code can be used as \HMAC in order to derive \hmac.
We note that the larger is the size of \hmac, the stronger privacy is 
preserved against the adversary \adversary, however, larger \hmac have diminishing
effect on the compression potential of the \cloud, as the \cloud can not compress \hmac efficiently.
We will discuss this in more detail in Section~\ref{sec:results}.
The \user then encrypts the \local using a symmetric encryption \encrypt with a secret key \enckey
 to generate \enclocal.
Unlike Dual Deduplication, where \local is stored locally in the \user,
the \user in \name only stores \enckey locally and sends the encrypted value
\enclocal to the \cloud.
As for the exact encryption algorithm, the \user is free to choose
the encryption algorithm and the size of the used private key.
This gives the flexibility to the \user to choose the best
encryption algorithm based on the use case, computational power
and desired privacy guarantees.
Finally, the user stores \hmac, and private keys for both \hmac and \enclocal,
and sends \hmac, \outsource, and \enclocal to the \cloud.

(3) The \cloud performs deduplication on the received \outsource, as in~\cite{sehat2021yggdrasil}.
We note that the \cloud would like to compress the received \enclocal in order to
provide better storage efficiency.
However, as the sizes of \enclocal is significantly small
compared to \outsource, and the ciphertext output by 
the symmetric encryption scheme looks random to the \cloud,
reducing the fingerprint of the received \enclocal is
not efficient and worthwhile.
%the \cloud achieves less than ideal compression rate for \enclocal and \hmac.

\subsection{Secure File Sharing}

This stage includes three parties, namely,
a \user that wants to share its data, denoted by \sender;
a \user that wishes to receive the data, i.e., \receiver;
and the \cloud.
We further assume that both \sender and \receiver use or have access
to the same PRNG, so that when using the same seed, the output of the
PRNG for both entities is exactly the same.
(4) When a \sender wants to share its data, it sends \hmac,
the private key for \HMAC, i.e., \hmackey and
the private key for \encrypt, i.e., \enckey
to the \receiver. (5) The \receiver uses \hmac, which is acting as
the file identifier to the data on the \cloud,
 to retrieve the respective \outsource and \enclocal from the \cloud.
 (6) After the \cloud receives the download request from the \receiver,
 it decompresses the data, by reversing the steps it took during the compression step,
 and sends the \outsource and \enclocal to the \receiver.
(7) As the receiver now has the components necessary to reconstruct the original file \file.
It follows the following steps to reconstruct the \file and checks the integrity of the data.
\begin{enumerate}
	\item (8) Decrypt \enclocal using the secret key \enckey to obtain the plaintext \local.
	\item (9) Use \local and \outsource in order to reconstruct \file by inserting the values stored in \local into
	their positions, generated by the seed stored in \local and the PRNG.
	\item (10) Calculate the \HMAC of \file and verify the result with \hmac to ensure the integrity of the data.
\end{enumerate}

The integrity check in the \receiver ensures that the file \file
is reconstructed successfully and is the same file \file that was originally in the \sender. 

% To allow compilation of the file
% !TeX root = ./../Bifrost.tex
\section{Privacy Analysis of \name}
\label{sec:privacyanalysis}
%\subsection{Compression ratio}
%
%We analyze the compression ratio in the \cloud, when using \name to store and share data between \users.
%As mentioned in section~\ref{sec:contribution}, the \cloud performs deduplication on the received \outsource.
%However, the received \enclocal and \hmac are stored without compression.
%Therefore, the required storage on the \cloud is equal to the required storage for all the \outsource,
%plus the required storage for the \hmac and \enclocal for each file that the \users has sent to the cloud.
%The required storage for the received \outsource heavily depends on the structure of the data 
%and how well the \cloud is able to compress the data.
%Our experiments show that in \name, the \cloud achieves between 68-92\% compression rate on \outsource.
%This amount of compression is competitive considering that the dual deduplication technique provides strong security
%against an adversary, which is not provided by current state-of-the-art compression techniques.
%%Furthermore, by using a compression method on top of the deduplication provided by \cite{sehat2021yggdrasil},
%%the \cloud achieves around 30\% compression rate for a dataset of HDFS files.
%
%The size of \hmac and \enclocal for each file is fixed and is dependent on the encryption and HMAC technique used.
%We discuss this in more detail in section~\ref{sec:results}, where we provide numerical results for some of the techniques used.

\begin{figure*}[!t]
	\centering
	\subfloat[$\sorg = 256$]{
		\begin{tikzpicture}
			[scale=0.8]
			\begin{axis}[
				xlabel= $\nodel$,
				ylabel=$\compratio$,
				xlabel style={font=\large},
				ymin=.8,
				ymax=1.5,
				grid=major,
				legend pos= north east]
				\addplot[mark=+, smooth ,black, thick] plot coordinates {
					(2,.9473)
					(4,.9308)
					(6,.9184)
					(8,.9116)
					(10,.9073)
					(12,.8694)%MIN
					(14,.8747)
					(16,.8821)
					(18,.8905)
					(20,.8994)
					(22,.9076)
					(24,.9123)
					(26,.9167)
				};
				\addlegendentry{$\hmackey=256, \enckey = 128$}
				
				\addplot[mark=x, smooth, dotted, red, thick, mark options={solid}] plot coordinates {
					(2,1.1832)
					(4,1.1722)
					(6,1.1591)
					(8,1.1507)
					(10,1.1238)
					(12,1.1021)%MIN
					(14,1.1133)
					(16,1.1262)
					(18,1.1297)
					(20,1.1373)
					(22,1.1467)
					(24,1.1517)
					(26,1.1562)
					
				};
				\addlegendentry{$\hmackey=512, \enckey = 128$}

				\addplot[mark =o, smooth,green, mark options={solid}] plot coordinates {
					(2,1.0022)
					(4,.9892)
					(6,.9762)
					(8,.9708)
					(10,.9662)
					(12,.9156)%MIN
					(14,.9321)
					(16,.9408)
					(18,.9492)
					(20,.9547)
					(22,.9634)
					(24,.9718)
					(26,.9792)
				};
				\addlegendentry{$\hmackey=256, \enckey = 256$}
				
				\addplot[mark = oplus, smooth, blue, thick] plot coordinates {
					(2,1.2441)
					(4,1.2330)
					(6,1.2184)
					(8,1.2114)
					(10,1.1881)
					(12,1.1643)%MIN
					(14,1.1751)
					(16,1.1871)
					(18,1.1876)
					(20,1.1981)
					(22,1.2052)
					(24,1.2121)
					(26,1.2183)
				};
				\addlegendentry{$\hmackey=512, \enckey = 256$}
				
			\end{axis}
	\end{tikzpicture}}
	\subfloat[k=8]{
		\begin{tikzpicture}
			[scale=0.8]
			\begin{axis}[
				xlabel= $\sorg$,
				ylabel=$\compratio$,
				xlabel style={font=\large},
				grid=major,
				legend pos= north east]
				
				\addplot[mark=+, smooth ,black, thick] plot coordinates {
					(64,1.2218)
					(100,1.0887)
					(128,0.9652)
					(160,0.9256)
					(200,0.8917)
					(256,0.8694)%MIN
					
				};
				\addlegendentry{$\hmackey=256, \enckey = 128$}
				
				\addplot[mark=x, smooth, dotted, red, thick, mark options={solid}] plot coordinates {
					(64,2.2158)
					(100,1.8717)
					(128,1.6738)
					(160,1.4256)
					(200,1.2432)
					(256,1.021)%MIN
					
				};
				\addlegendentry{$\hmackey=512, \enckey = 128$}

				\addplot[mark =o, smooth,green, mark options={solid}] plot coordinates {
					(64,1.4816)
					(100,1.2087)
					(128,1.1254)
					(160,1.0648)
					(200,0.9743)
					(256,0.9156)%MIN
				};
				\addlegendentry{$\hmackey=256, \enckey = 256$}
				
				\addplot[mark = oplus, smooth, blue, thick] plot coordinates {
					(64,2.5282)
					(100,2.1443)
					(128,1.9132)
					(160,1.6626)
					(200,1.4513)
					(256,1.2183)%MIN
				};
				\addlegendentry{$\hmackey=512, \enckey = 256$}
				
			\end{axis}
	\end{tikzpicture} }
	
	\caption{The compression rate for different techniques }
	\label{fig:res1}
\end{figure*}
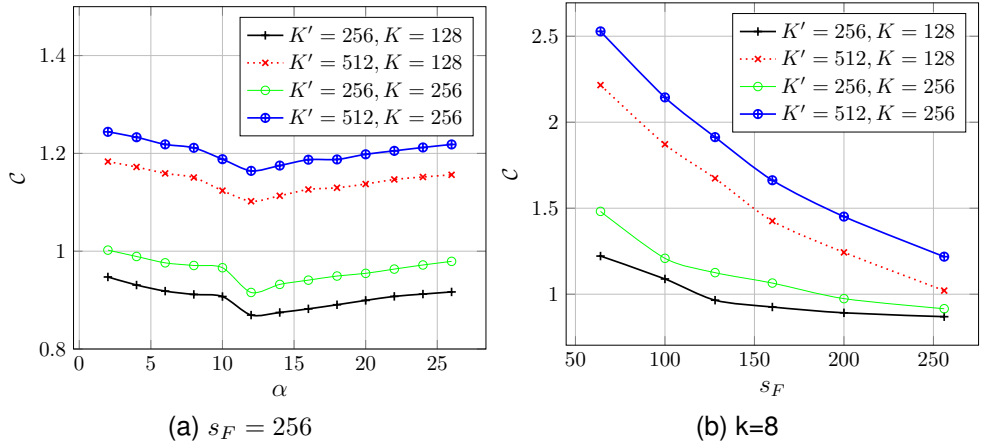

In \name we use three main cryptographic primitives on the \users' side: 
a \PRNG (used in \transformation), a symmetric encryption scheme \encrypt, and an \HMAC. 
Intuitively, we protect data privacy from a malicious \cloud in two steps: 
first we delete a portion of the original file (in \transformation,  the indexes of the values to be deleted are determined by the output of a \PRNG on a given seed), 
then we encrypt deviation \deviation (which contains the seed and the deleted values) using a secure symmetric encryption scheme \encrypt. 
Notably, \outsource is a punctured version of \file, and  
this transformation has been shown to provide strong privacy guarantees as it mimics a deletion channel~\cite{sehat2021yggdrasil}. 
In our security analysis, we idealize \PRNG to a random oracle, and demand \encrypt to be IND-CCA2 secure. 
Finally, to ensure the integrity of the outsourced data, we employ a MAC tag \hmac, generated by \HMAC on the plaintext \file. 
Although \HMAC is designed to provide data integrity only, in our security analysis we model it as a random oracle, which ensures \hmac leaks no information about the plaintext data. 

In order to argue the privacy of \bifrost we make the following set of assumptions on the 
knowledge and the computational power of our adversary \adversary.

\begin{assumption}\label{ass:distribution}
	The adversary \adversary knows the probability distribution of the plaintext data \file, and has access to all resourced uploaded to the \cloud, i.e., \outsource, \enclocal, and \hmac.  
\end{assumption}
\begin{assumption}\label{ass:keys}
	The adversary \adversary has no access to \users' cryptographic key material (in particular, \enckey and \hmackey are unknown to \adversary). 
\end{assumption}
\begin{assumption}
	The adversary \adversary has computational power bounded by polynomial-time algorithms, executed on a classical computer (we do not consider quantum adversaries).  
\end{assumption}
Furthermore, we make the following assumptions on the cryptographic primitives used in \bifrost:
%\begin{assumption}
%	The \cloud does not have knowledge about the position and the value of the deleted symbols in \transformation.
%\end{assumption}
\begin{assumption}\label{ass:rom}
	The algorithms \PRNG and \HMAC act as random oracles; that is: their output are random strings uncorrelated to the input, except for consistency, i.e., the same input returns the same output.
\end{assumption}
\begin{assumption}\label{ass:ind-cca2}
	The \encrypt algorithm satisfies the security notion of indistinguishablility under chosen plaintext attack (IND-CCA2).
\end{assumption}
\begin{assumption}\label{ass:unforgeable}
	The \HMAC algorithm is unforgeable.
\end{assumption}
%\begin{assumption}
%	\HMAC has the properties of a random oracle, i.e., the output of the \HMAC is random and uncorrelated to the input, except for consistency, i.e., the same input returns the same output.
%\end{assumption}

\noindent
Assumptions \ref{ass:rom} and \ref{ass:ind-cca2} have the two following major implications: 

First, given a string \outsource, \adversary cannot identify the position of the symbols deleted by \transformation better than a random guess. 
In particular, modelling the \PRNG as a random oracle makes \transformation behave like a deletion channel on \file. It is proven that the deletion channel is a hard problem for the receiver (\adversary in this case) who cannot deterministically reconstruct the original string using the received punctured version~\cite{mitzenmacher2009}.

Second, given, \outsource, \enclocal, and \hmac, it is computationally infeasible for \adversary to retrieve values deleted by \transformation and recover the original plaintext file \file. This is essentially due to the fact that \enclocal and \hmac are output of random oracles and \outsource is the output of a deletion channel (as discussed in the previous point). 

%If \adversary cannot break the PRNG, the output of the \transformation function looks like the output of a deletion channel.
%It is proven that the deletion channel is a hard problem for the receiver and the receiver cannot construct the original string by receiving the output of a deletion channel~\cite{mitzenmacher2009}.
%Therefore, assumption 6 follows directly from assumption 5.
%Based on assumption 1, we conclude that the \cloud learns no information about the file \file
%by receiving the outsource \outsource, as the concept of the deletion channel provides information-theoretic security.
%
%The \cloud receives \hmac, \enclocal and \outsource from the \user.
%Based on the assumptions 1, 2 and 4, neither of these pieces
%leak any information about the original file \file.
%Therefore, a malicious adversary \adversary does not gain any information 
%by receiving the outsourced data from the \user.
%We argue that if a malicious \user gains access to the data stored on the \cloud,
%It can not learn more knowledge than the \cloud, 
%proving the privacy of the data stored on the \cloud against this type of adversary.

We move on to proving integrity of the downloaded data.
In this case we are concerned with an adversary that acts as a malicious \cloud, modifies the outsourced data, and returns inconsitent values to download requests. 
To detect and deter such missbehaviour, \bifrost employs a \HMAC. 
We recall that in \bifrost, the tag \hmac output by \HMAC serves both as a file identifier and as token for integrity check. 
When the \user (\sender) shares its data with another \user (\receiver), it sends \hmac on a P2P, private channel. 
The \receiver uses \hmac to check the validity of the data it receives from the \cloud.
By Assumptions \ref{ass:keys} and \ref{ass:unforgeable} , \adversary has no access to the secret key \hmackey and \HMAC is unforgeable. Therefore, it is computationally infeasible for \adversary to modify \outsource or \enclocal without being detected. 
%Besed on assumption 3, 
%the \cloud is not able to calculate a collision for a given value \hmac,
%therefore, the receiver is able to check if the received \outsource form the \cloud 
%is the correct data,
%therefore, the malicious \cloud is not able to alter the stored data, without being noticed. 

% To allow compilation of the file
% !TeX root = ./../Bifrost.tex

\section{Numerical Results and Discussion}\label{sec:results}

In this Section, we provide numerical results gathered by simulation of \name.
We use a C++ based implementation consisting of a cloud and 2 clients, 
where one client stores data on the cloud and then shares it with the second client.
We use 10GB of HDFS and 8GB of DVI data-sets. For the sake of having more diverse data,
we use a random portion of both datasets in our experiments.
We set the size of each original string \file to 2048 bits and a symbol size of 8 bits.
In other words, our data consists of strings of 256Bytes, and the deletion in the client as well as the
deduplication on the cloud is performed on the Byte level.
For our PRNG, we use a seed size of 32 bits.

In order to analyze the trade-off between security and the compression rate/transmission overhead
in \name, we experiment with multiple setups as \HMAC and encryption function (\encrypt).
For our encryption function, we use AES with key sizes of 128 bits and 256 bits.

For our HMAC, we used two different setups,
\begin{enumerate*}
	\item HMAC based on SHA-2 with a key size of 256 bits, resulting in a hash size of 256 bits.
	\item  HMAC based on SHA-2 with a key size of 512 bits, resulting in a hash size of 512 bits.
\end{enumerate*}

Note that the larger size of keys in HMAC and encryption,
i.e., the larger size of \hmackey and \enckey,
the system would provide higher security for the data, as the encryption and the HMAC would be 
harder to break for the adversary. However, we note that all of the setups used in this simulation are 
currently considered computationally secure in real-world applications.

\begin{table*}[!t]
	\caption{The transmission overhead in multiple secure file sharing systems.}
	\vspace{-1.3em}
	\begin{center}
		\renewcommand{\arraystretch}{1.3} %makes table rows more spacious
		{{	\begin{tabular}{p{0.1\textwidth} >{\RaggedLeft}p{0.15\textwidth} 
						>{\RaggedLeft}p{0.15\textwidth}
						>{\RaggedLeft}p{0.05\textwidth}
						>{\RaggedLeft}p{0.05\textwidth}}
					%					\hline			
					Protocol name & \cloud to \user & \user to \user  & 
					Privacy & 
					Compression \\
					\hline
%					& HDFS  & DVI & HDFS & DVI & & & \vspace{.2em}\\
%					%\hline
%					%\hline
					\rowcolor{gray!20} %to color rows
					\name 
					& $402MB$ & $2372 bits$ & Yes & Yes\\
									%\hline
					Plutus~\cite{plutus} 
							& $404MB$ & $1024 bits$  & Yes & No\\
							
%					GD+MKRE~\cite{mkre} 
%					& $ 43.20\%$ &  $29.16\%$ &$ 12.8\%$ &  $8.52\%$ & YES & Random\phantom{bb} Oracle & Honest but Curious\\
					%\hline
					\rowcolor{gray!20} %to color rows
						SEGShare~\cite{fuhry2020segshare} 
					& $404MB$ & $2200bits$ & Yes & No\\

					No Cloud 
				 	& $0$ & $400MB$ & Yes & No \\
%					%
	%				\hline
					\rowcolor{gray!20}
						Only Cloud 
					 & $ 400MB$ & 0 & No & Yes \\
%					%					\hline
				\end{tabular}
			}
		}
		\label{table}
	\end{center}
	\vspace{-2em}
\end{table*}	

\subsection{Compression rate}\label{sec:resultsA}

In Fig.~\ref{fig:res1}, we show the compression rate on the cloud for different setups of the HMAC and
encryption. 
For this experiment, we have considered chunk size of up to 256 Bytes, which is the upper bound for current
dual deduplication system as mentioned in \cite{sehat2021yggdrasil}.
In Fig.~\ref{fig:res1}.a, we calculate the compression rate for different values of $\nodel$ when the chunk size
\sorg = 256 Bytes.
As this Figure suggests, for lower values of $\nodel$, the more deletions we have in the client, i.e., higher the number of \nodel is,
the more compression we achieve on the \cloud.
This is due to two facts. 1. For lower number of deletions, the size of the secret piece \local is lower than the size of the key used for encryption, therefore, the AES coding protocol uses padding to generate blocks of 128 bits, producing overhead in the \cloud; and 2. By having more deletions, the size of the outsource $\sbase$ is reduced, increasing the potential for deduplication in the \cloud.
After a certain threshold, this trend is reversed and the \cloud loses its compression potential by increasing the value of $\nodel$. This behavior occurs as by having more deletions, the size of \local grows, which is the part of the data that is encrypted and therefore cannot be efficiently compressed.
We believe that this threshold depends on the encryption algorithm.
For this particular experiment the optimal compression rate is for when
$\nodel = 12$, i.e., when the size of \local is exactly 128 bits,
which is exactly the same value as the cipher blocks used in AES.

In Fig.~\ref{fig:res1}.b, we calculate the best compression rate for different values of \sorg.
In this plot, we have illustrated the best compression rate achievable for each value of \sorg,
which is when $\nodel = 12$. As this graph suggests, we achieve the best compression rate
for the highest value of \sorg, as in this case, the proportion of \enclocal and \hmac to
the total size of the the database is lower.

In our particular setup, we are able to achieve a compression rate of $86.94\%$, of which
$17.46\%$ is the size of \hmac and \enclocal and $69.48\%$ is the size of the \outsource.
This compression rate is achieved for the lowest values on the size of the keys \hmackey and \enckey.
As an alternative approach, we can encrypt the \local generated by multiple chunks in a single file using the 
same \enckey and produce a single \hmac for multiple chunks in the file.
Using this approach, we reduce the required storage for the \hmac on the \cloud,
as well as the required storage on the \user for the keys \hmackey and \enckey.
However, this approach reduces the security of \name, as if an adversary is able to break a single chunk
and obtain the \enckey, it can break all the chunks that used the same key in \encrypt. 
%We note that the compression rate is likely to decrease even further for higher values of \sorg,

\subsection{Transmission overhead}

For this paper, we have analyzed the theoretical overhead on the transmission between \cloud and \users in \name,
and compared it with the extreme cases of file sharing as well as some secure file sharing systems.
Table~\ref{table} shows the required transmission between different parties when using \name, Plutus~\cite{plutus}, SEGShare~\cite{fuhry2020segshare} and the extreme cases of only using the \cloud and only using the P2P channel.
Note that these values are theoretical and are for a file size of $\sorg = 200MB$.
%As we noted in section~\ref{sec:resultsA}, the performance of \name increases for higher values of \sorg,
in this table, we assumed that the \local for all chunks of the file is encrypted with a single key \enclocal, with the size of 128bits;
 and we produce a single \hmac for every 100 chunks of the file using a unique key \hmackey with a size of 256 bits.
 This leads to the total transmission overhead of $384$ bits for the keys and $2MB$ for the \hmac in the P2P channel.
 For the transmission in the channel between the \cloud and the \receiver, we note that the \receiver is required to download the whole \outsource, as the \cloud decompresses the data before sending it to the \receiver.
 Therefore, the required transmission in that channel is equal to the size of the file and the file identifier (\sorg + \sizefid).
 Comparing these schemes, we see that \name achieves a competitive transmission cost in the case of file sharing between two \users, while having a high compression rate on the \cloud, which is not provided in the other secure file sharing systems.
 
%\input{sections/rel_work}
% To allow compilation of the file
% !TeX root = ./../Bifrost.tex
%\vspace{-5pt}
\section{Conclusion and Future Work}\label{sec:conclusion}

In this work, we propose \name, an innovative secure file sharing system
based on dual deduplication~\cite{sehat2021yggdrasil}, which provides computational privacy for the data of
the clients against a malicious cloud or client, while allowing the
untrusted cloud storage to compress the data.
 % to around 90\% of its original size.
We achieve this goal by storing a fixed amount of data for each file in the clients,
regardless of the size of the file,
making the proposed solution scalable to large amounts of data.
Our analysis shows that the computational security is provided by storing only 650 bits
for each file in the client side.

The proposed method is unique in its focus on provide compression on the data, allowing
for more clients to use the cloud storage system to store and share their data in a secure way.
Our experiments show that the cloud can achieve 86.9\% compression rate, without reducing
the security of \name.
Future work will focus on (a) revoking the rights of other clients to read or write the data stored on the cloud,
(b) analysis of the trade-off between security and compression for different security and compression schemes, and
(c) protection against malicious receiver without revealing unwanted information.
As dual deduplication is a recent topic, the privacy and security opportunities of it will require further work.
%We note as the aspect of \name and the whole idea of Dual Deduplication is quite new, there are multiple aspects of privacy
%and security that need to be addressed in future work.

\section{Acknowledgment}
This work was partially financed by the SCALE-IoT project (DFF-7026-00042B) granted by the Danish Council for Independent Research, the Innovation Fund Denmark (8057-00059B), the AUFF Starting Grant AUFF2017-FLS-7-1, Aarhus University’s DIGIT Centre and by the Swedish Foundation for Strategic Research (RIT17-0035), and the strategic research area ELLIIT. 

\bibliographystyle{unsrt}
\bibliography{references.bib}

\end{document}